# Realizing spin-dependent gauge field with biaxial metamaterials

Fu Liu[1+], Tao Xu[2+], Saisai Wang[2], Zhi Hong Hang[2*] and Jensen Li[1*]

[1] School of Physics and Astronomy, University of Birmingham, Birmingham B15 2TT, United Kingdom

[2] College of Physics, Optoelectronics and Energy & Collaborative Innovation Center of Suzhou Nano Science and Technology, Soochow University, Suzhou 215006, China

+ These authors contributed equally to this work

**Artificial magnetic field in electromagnetism is becoming an emerging way as a robust control of light based on its geometric and topological nature. Other than demonstrating topological photonics properties in the diffractive regime using photonic crystals or arrays of waveguides, it will be of great interest if similar manipulations can be done simply in the long wavelength limit, in which only a few optical parameters can be used to describe the system, making the future optical component design much easier. Here, by designing and fabricating a metamaterial with split dispersion surface, we provide a straight-forward experimental realization of spin-dependent gauge field in the real space using a biaxial material. A "magnetic force bending" for light of desired pseudospins is visualized experimentally by such a gauge field as a manifestation of optical spin Hall effect. Such a demonstration is potentially useful to develop pseudospin optics, topological components and spin-enabled transformation optical devices.**

The concept of a photonic gauge field and its associated artificial magnetic field is an emerging way to manipulate light by exploring the geometrical or topological properties of light [1-11]. More robust operations or more intuitive design principles can thus be triggered comparing to conventional approaches. Currently, it has become central to various optical devices from geometric-phase metasurfaces for beam structuring , when the concept is applied to the polarization space, to topological photonics for generating one-way edge states, when the concept is applied to the reciprocal space [12,13]. For the case of a photonic gauge field in the real space, it can be regarded as the analogy of a vector potential acting on electron motion, whose quantum-mechanical effect is historically considered as the Aharonov–Bohm effect.

While a light is bent conventionally through a spatially varying profile of dispersion surfaces with varying sizes or shapes, a photonic gauge field in the real space essentially bends light through the shifting of the local dispersion surfaces [14-17]. It enables us to demonstrate various peculiar wave phenomena such as negative refraction, gauge-field enabled waveguiding, one-way edge state and asymmetric wave transport [1-11]. Real-space photonic gauge field realizations are proposed, while most of them are in a network of resonators or waveguides with direction-dependent coupling, achievable by different optical paths [1-3], dynamic modulation [4-9], magneto-optical effect [10], or an inhomogeneously strained structural profile [11]. If a real-space photonic gauge field can be "materialized" to work in the effective medium regime and a few effective medium parameters are enough to capture its physical description, we shall be able to apply the corresponding design principle to a simpler model to promote the development of more compact optical components. In fact, the materialization approach has been applied to design topological metamaterials, which is relevant in considering gauge field and artificial magnetic field in the reciprocal space [18-20]. On the other hand, a tilted anisotropic medium, by considering a field transformation approach [21-23], is theoretically suggested to generate a real-space photonic gauge field, but the stringent requirement of matched magnetic and electric response complicates experimental realization.

Here, we experimentally demonstrate a real space photonic gauge field using a biaxial medium with only electric response. We simply refer it as a gauge-field medium. In the current scheme, as long as the different permittivities along the three principal axes satisfy a geometric-mean relationship, a gauge field exists for waves traveling on a particular chosen working plane. This gauge-field pushes the originally degenerated circular dispersion surfaces on the working plane towards opposite directions, depending on which decoupled polarization, or called pseudospin, is employed. Pseudospin-dependent beam splitting and propagation near the associated singular (diabolical) points are demonstrated. Our work provides a feasible route to incorporate gauge field using metamaterials. Furthermore, there is actually no fundamental limitation to push the current working principle to the optical regime as the required mean relationship is discussed. Such a biaxial medium approach may also allow future generalization to more complicated gauge-field media, for example, to realize a non-Abelian gauge field using biaxial media [24].

We start from the gauge field medium, which is represented by permittivity and permeability tensors:

$$\bar{\bar{\epsilon}} = \begin{bmatrix} n_0 & 0 & A_y \\ 0 & n_0 & -A_x \\ A_y & -A_x & n_z \end{bmatrix}, \quad \bar{\bar{\mu}} = \begin{bmatrix} n_0 & 0 & -A_y \\ 0 & n_0 & A_x \\ -A_y & A_x & n_z \end{bmatrix}, \quad (1)$$

where $n_0$ and $n_z$ are the common principal values and the U(1) gauge field $\boldsymbol{A} = A_x\hat{x} + A_y\hat{y}$ is the tilted anisotropy terms with the same magnitude but opposite signs in the two tensors. It is defined for wave propagation on the $x - y$ plane, satisfying a wave equation using $E_z$ and $H_z$ as basis [12]. The eigenmodes of the medium are governed by the secular equation

$$\begin{bmatrix} k^2 - k_0^2(n_0 n_z - A^2) & -2ik_0\boldsymbol{A}\cdot\boldsymbol{k} \\ 2ik_0\boldsymbol{A}\cdot\boldsymbol{k} & k^2 - k_0^2(n_0 n_z - A^2) \end{bmatrix} \begin{bmatrix} E_z \\ iH_z \end{bmatrix} = 0,$$

where $\boldsymbol{k} = k_x\hat{x} + k_y\hat{y}$ is the wavevector, $k_0$ is the wavenumber and $A = |\boldsymbol{A}|$. Two shifted circular dispersions are attained. Each of them corresponds to a constant and decoupled eigen-polarization with either $E_z = H_z$, defined as pseudo spin-up $\psi_+$, or $E_z = -H_z$, defined as pseudo spin-down $\psi_-$. The size of the shifting of the dispersion surfaces in the reciprocal space is given by $\mp k_0 A$ for the two pseudo-spins (Fig. 1).

The requirement of responses in matched values and the additional tilted anisotropy terms in both tensors are relaxed in two steps. The so-called reduced parameter approximation technique is often employed for metamaterials requiring both electric and magnetic responses [25], which is valid when impedance mismatch is negligible or when the material profile is slowly varying in the scale of wavelength. First, by exploiting the fact that the information of the in-plane fields is lost in establishing the secular equation, one can prove that the following medium

$$\bar{\bar{\epsilon}} = \begin{bmatrix} n_0 & 0 & 2A_y \\ 0 & n_0 & -2A_x \\ 2A_y & -2A_x & n_z + 3A^2/n_0 \end{bmatrix}, \quad \bar{\bar{\mu}} = \begin{bmatrix} n_0 & 0 & 0 \\ 0 & n_0 & 0 \\ 0 & 0 & n_z - A^2/n_0 \end{bmatrix}, \quad (2)$$

gives exactly the same secular matrix, hence the same dispersion surfaces (shape and size) and constant polarizations for the two modes. It allows us to lump the tilted anisotropy terms from $\bar{\bar{\mu}}$ to $\bar{\bar{\epsilon}}$ only. Second, we confine to a particular case $n_z = n_0 + A^2/n_0$, making $\bar{\bar{\mu}}$ to be isotropic. A conventional reduced parameter approximation is further applied by dividing $\bar{\bar{\mu}}$ and multiplying $\bar{\bar{\epsilon}}$ by $n_0$ at the same time. The final medium becomes

$$\bar{\bar{\epsilon}} = \begin{bmatrix} n_0^2 & 0 & 2n_0 A_y \\ 0 & n_0^2 & -2n_0 A_x \\ 2n_0 A_y & -2n_0 A_x & n_0^2 + 4A^2 \end{bmatrix}, \quad (3)$$

with $\bar{\bar{\mu}}$ being the identity matrix (i.e. simply the free-space).

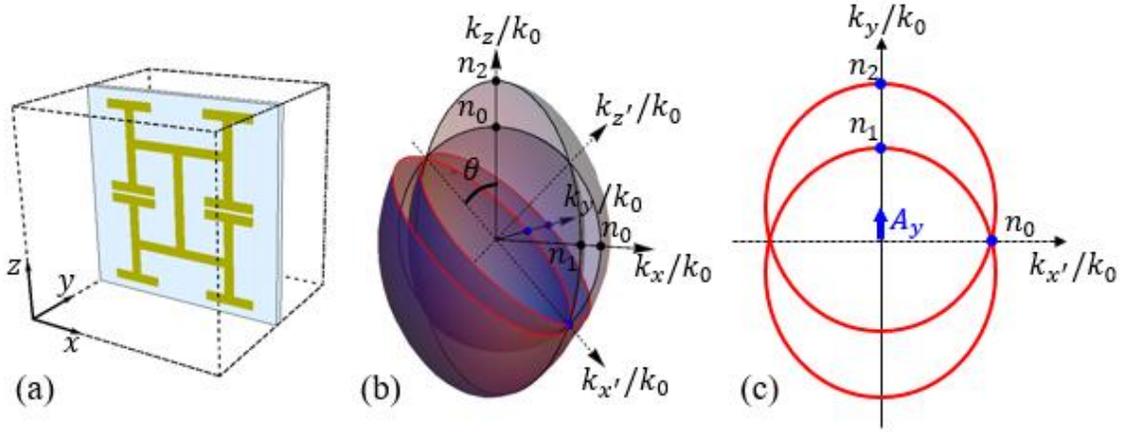

**Fig. 1 Obtaining gauge field from biaxial dielectric medium. a**, Cubic lattice with unit cell design containing a two-level fractal structure. Principal axes are along the Cartesian axes. **b**, 3D dispersion surface. Only three-quarters of it is shown with cutting plane on the $k_x - k_z$ plane and the $k_{x'} - k_y$ plane. The direction of $k_{x'}$-axis is defined to pass through the degeneracy point between the circle of index $n_0 = \sqrt{\epsilon_y}$ and the ellipse with principal axes indices $n_1 = \sqrt{\epsilon_z}$, and $n_2 = \sqrt{\epsilon_x}$ on the $k_x - k_z$ plane. **c,** Dispersion surface on the $k_{x'} - k_y$ plane ($k_{z'} = 0$). These are two shifted ellipses with opposite shifting by a gauge field of $\boldsymbol{A} = A_y \hat{y}$.

We recognize that the resultant medium in Eq. (3) is simply a biaxial dielectric medium. If we write $\boldsymbol{A} = A\hat{n}$ where $\hat{n}$ is a unit vector lying on the $x$-$y$ plane, a clockwise rotation of the medium by an angle $\theta = \mathrm{acot}(A/n_0)/2$ about $\hat{n}$ can diagonalize $\bar{\bar{\epsilon}}$ to its principal values $\epsilon_n = n_0^2$, $\epsilon_u = n_0^2 \tan^2\theta$ and $\epsilon_v = n_0^2 \cot^2\theta$ along $\hat{n}$, $\hat{z}\times\hat{n}$ and $\hat{z}$ respectively. In other words, suppose we have a biaxial dielectric medium $\bar{\bar{\epsilon}} = \epsilon_n \hat{n}\hat{n} + \epsilon_u \hat{u}\hat{u} + \epsilon_v \hat{v}\hat{v}$ satisfying a geometric-mean relationship on its permittivities along the three orthogonal directions:

$$\epsilon_n^2 = \epsilon_u \epsilon_v. \tag{4}$$

We reach a simple recipe to generate a gauge field

$$\boldsymbol{A} = \pm \hat{n}(\sqrt{\epsilon_v} - \sqrt{\epsilon_u})/2 \tag{5}$$

in such a medium, for waves propagating on the $x$-$y$ plane, which is now defined as the plane containing $\hat{n}$ and $\hat{u}\cos\theta - \hat{v}\sin\theta$. The "$\pm$" sign corresponds to the two choices of $\theta$ with the same magnitude but opposite signs.

The biaxial medium satisfying Eq. (4) has the same dispersion surfaces of the ideal gauge field medium and the decoupled pseudospins are only subject to a redefinition to $n_0 E_z = \pm H_z$. Resultant from the criteria, the polarization remains constant and ensures that when the wave changes direction in the medium, the two pseudospins stay decoupled with each other.

We adopt a biaxial metamaterial for realization, with a unit cell shown in Fig. 1(a). Printed circuit boards (PCBs) with a square array of metallic fractal structures are periodically stacked along the $y$ direction to construct a three-dimensional (3D) metamaterial. We assume $\epsilon_x > \epsilon_y > \epsilon_z$. Figure 1(b) shows a typical dispersion surface of such a biaxial metamaterial in the 3D reciprocal space. For clarity, the cross-section on the $k_x - k_z$ plane is shown with thick black lines. For the polarization $(E_y, H_z, H_x)$, a circular dispersion surface of radius $n_0 = \sqrt{\epsilon_y}$ is expected due to the lack of magnetic response. There is also an ellipse with semi- minor / major axis of index $n_1 = \sqrt{\epsilon_z}$ /$n_2 = \sqrt{\epsilon_x}$ for the polarization $(H_y, E_x, E_z)$. The two dispersion surfaces intersect with each other at 4 diabolical points. A plane containing the $y$-axis can then be drawn to pass through a pair of these diabolical points, defined as the $k_{x'} - k_y$, as shown in Fig. 1(b). The $x' - y$ plane is our working plane for in-plane wave propagation. The dispersion surfaces on the working plane are zoomed in as solid red curves with the principal indices indicated in Fig. 1(c). They look like two split circles, suggesting a gauge field $\boldsymbol{A}$ can be defined to indicate the size of splitting, which is the same as in Eq. (5), now with $\epsilon_n = \epsilon_y$, $\epsilon_v = \epsilon_x$, $\epsilon_u = \epsilon_z$ and $\hat{n} = \hat{y}$. The "$\pm$" sign corresponds to the two choices of the plane to pass through the diabolical points. In the current biaxial metamaterial with arbitrary principal values, the working plane has an orientation (rotated by $\theta$ from the z-axis) governed by

$$\tan^2 \theta = \frac{\epsilon_x}{\epsilon_z} \frac{\epsilon_y - \epsilon_z}{\epsilon_x - \epsilon_y}, \tag{6}$$

where the geometric-mean criteria with $\epsilon_y = \epsilon_z \cot^2 \theta = \epsilon_x \tan^2 \theta$ is covered but not limited to. It indicates that the gauge field medium derived in previous section is not a singular case but rather a case that can be approached smoothly using the above consideration of biaxial medium. This adds tolerance in favor to experimental realization. A small deviation from the geometric-mean criteria only leads to a less circular dispersion surfaces and the a less independent polarization on the reciprocal space. Therefore, for a biaxial medium where the difference between the principal permittivities are not too big, we are expecting a broad working regime for the gauge field, as long as one of the principal permittivities is situated roughly in the middle between the other two principal values. In such a case, $\theta$ is around $\pm 45\,^\circ$.

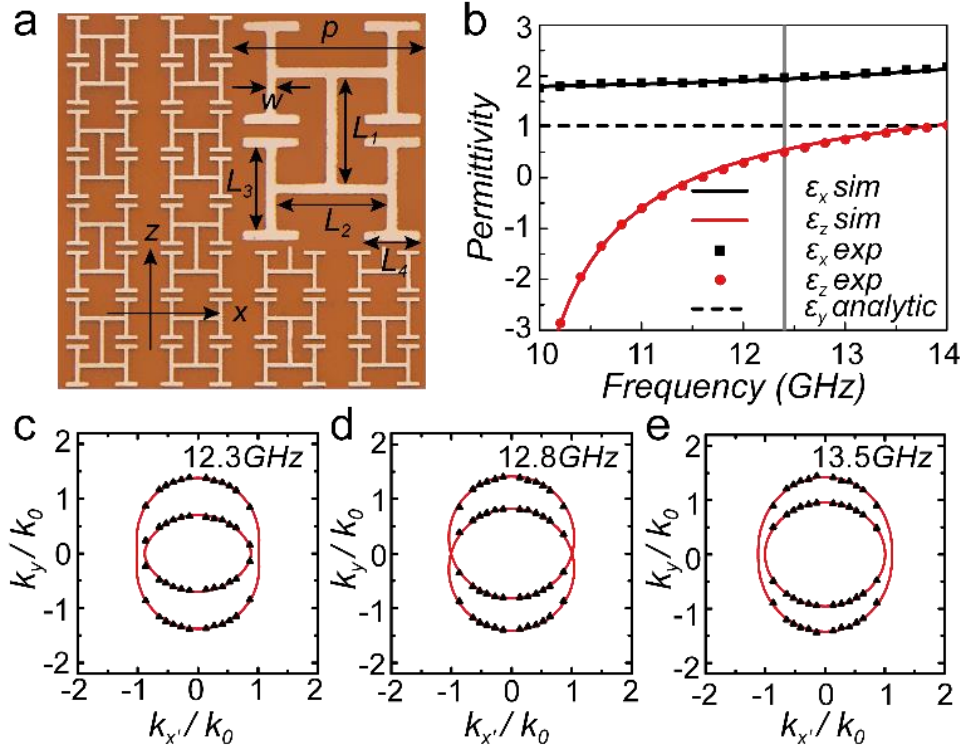

**Fig. 2 Spin-split circular dispersion surfaces. a,** Fabricated metamaterial to realize the gauge field. Copper fractal structure printed on PCB with geometry parameters $L_1 = 2.3$ mm $l_2 = 2.4$ mm $L_3 = 1.8$ mm $L_4 = 1.2$ mm, and copper width $w = 0.2$ mm. Copper layer has a thickness of 0.033 mm. The unit cell has a periodicity $p = 5$ mm in all directions. **b,** Effective medium $\epsilon_x$ and $\epsilon_z$ retrieved from simulation (solid lines) and experiment (symbols). The dashed line is $\epsilon_y$ analytically calculated by averaging the permittivity in $y$-direction. The vertical gray line indicates the working frequency 12.4 GHz at which the permittivity satisfies Eq. (4). **c-e,** Two dimensional dispersion surfaces at 12.3, 12.8, and 13.5 GHz respectively on the $k_{x'} - k_y$ plane. Symbols are from experiment measurement while the red curves are the analytic dispersion from the effective medium. At 12.8 GHz, the dispersion surfaces for the two polarizations are degenerated at a point on the $k_{x'}$ axis while gaps are opened at the other two frequencies.

Based on this analysis, we move on to our fabricated biaxial metamaterial working in the microwave regime, as shown in Fig. 2(a). The inset shows one unit cell with periodicity $p = 5mm$ (in all three directions) and the geometric parameters with values specified in figure caption. Copper fractal structures are printed on FR4 PCB with thickness $0.115mm$ and relative permittivity 3.3. The permittivity $\epsilon_y$ is very close to one (1.016 after averaging, dashed line in Fig. 2(b)) due to a much smaller thickness of PCB than $p$. On the other hand, $\epsilon_x > 1 > \epsilon_z$ is achieved by inducing electric resonances of the fractal structure. Scattering S-parameters are measured for microwaves, with polarization in either $x$ or $z$ direction, impinging the sample at normal incidence. $\epsilon_x$ / $\epsilon_z$ is then retrieved from the S-parameters and is plotted as solid black / red symbols in Fig. 2(b), with excellent agreement with corresponding retrieved results from full-wave simulations plotted in solid lines. Due to a resonating mode below 10 GHz, $\epsilon_z$ is between 0 and 1 from 11.5 to 14 GHz. For $\epsilon_x$, as its resonance is much higher than 14GHz, its value stays around 2 from 10 to 14 GHz. The frequency dispersion embedded in our fabricated biaxial metamaterial actually helps to scan and pick the working frequency to satisfy the mean criteria. The working frequency with material parameters matching Eq. (4) is found at around 12.3 GHz, indicated by a vertical gray line in Fig. 2(b), where $\epsilon_x = 1.965$ and $\epsilon_z = 0.504$.

A key feature of a gauge field is its ability to shift dispersion surfaces. Here, we measure the dispersion surfaces on the $k_{x\prime} - k_y$ plane. To have generality among various frequencies for comparison, the plane is set at $\theta = 45°$ as an approximation discussed above. The sample is now rotated by 45° in the *x-y* plane. The S-parameters at different incident angles are then measured. The wavevector stays on the $k_{x\prime} - k_y$ plane, equivalently the $x' - y$ plane in the laboratory frame. Since the near-field coupling between neighboring layers is negligible for our structure, only one layer of the fractal structure is employed to avoid the necessity of index branch selection from S-parameter retrieval. The results are shown in Fig. 2(c)-(e), for three slightly different frequencies 12.3, 12.8, and 13.5 GHz in order to follow gap closing and opening. The dispersion surfaces obtained from the retrieved effective material parameters at normal incidence (Fig. 2(b)) are plotted in red color where a good agreement can be found to experimental results at various incident angles indicated by symbols. From the results, two shifted circles with diabolical points at $k_y = 0$ are obtained at 12.8 GHz, which is slightly shifted from the predicted 12.3 GHz due to the deviation of $\theta$ from its ideal value. It can also be understood by plugging 45° into Eq. (6), which suggests the geometric-mean criteria should be modified to a harmonic-mean version: $1/\epsilon_n = (1/\epsilon_u + 1/\epsilon_v)/2$. As long as the differences in the principal permittivities are small, the various ways to take mean value do not differ significantly from each other and we expect the gauge field should still work with physics driven by split circular dispersions and decoupled pseudo-spin propagations.

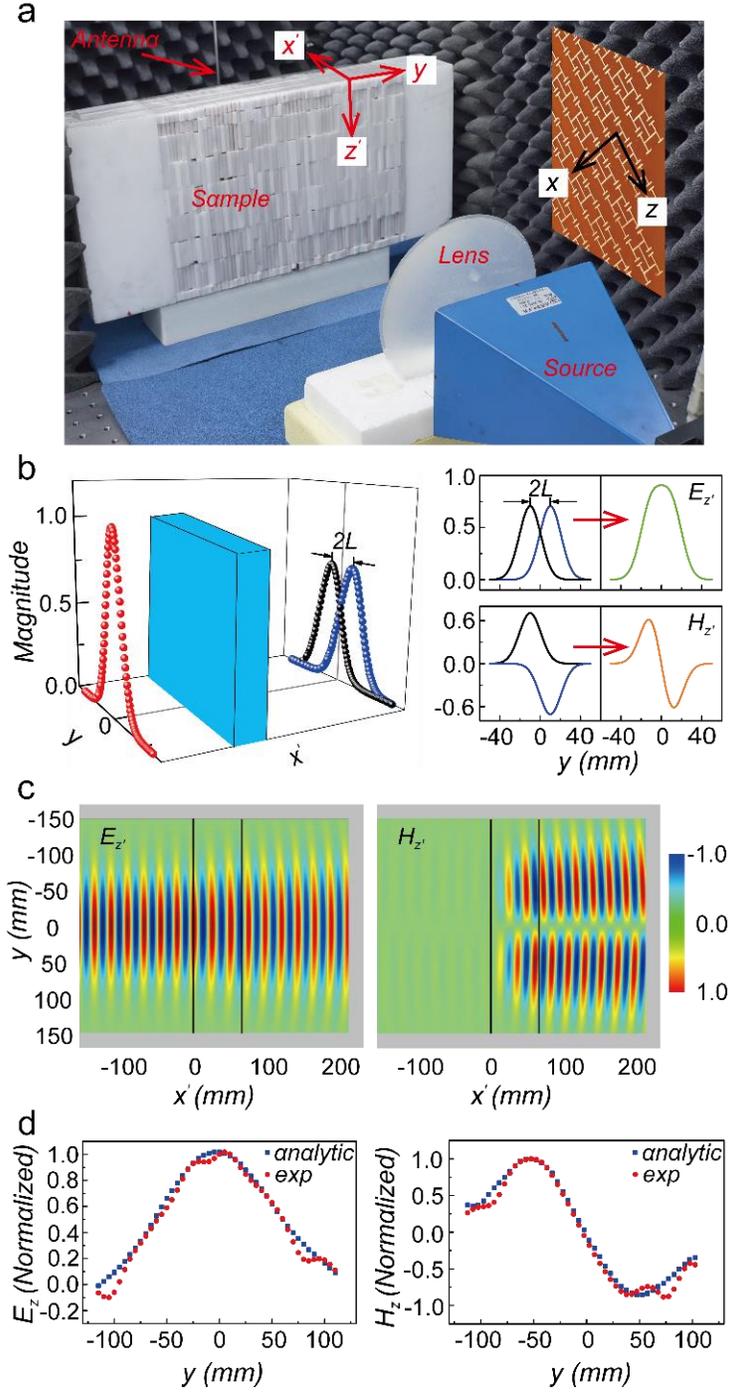

**Fig 3 Experiment Beam shift measurement. a,** experimental setup. **b,** Schematic with linear $E_{z'}$ polarization incidence. Beam splitting occurs for two pseudospins (black and blue) and the abrupt artificial magnetic field will induce a $L$ beam shifting of both spins. **c,** simulated $E_{z'}$ and $H_{z'}$ field distribution with $E_{z'}$ Gaussian beam incidence from left. **d,** The measured $E_{z'}$ and $H_{z'}$ field profile 30 mm away from the exit surface. $L$=17.5 mm beam shifting is found to be of best fit to analytically obtained profiles.

A gauge field medium bends photon by shifting dispersion surfaces. In a gradient material profile, it bends photon using an artificial magnetic field, which relates to the gauge field $\boldsymbol{A}$ by $\boldsymbol{B} = \nabla \times \boldsymbol{A}$. When we have a step change of gauge field $\boldsymbol{A}$ in this work, such a "magnetic force" bending can be understood as a surface effect. Similar interpretation applies to a pseudo-electric field from gradient index $\boldsymbol{E} \sim \nabla n$. If the refractive index is gradually changing, there is a layer with bulk pseudo-electric field, an incident beam will change direction by the pseudo-electric field, with a lateral shift across the interface. For a step change of index, all these effects are lumped to the surface, there is no lateral shift but a change of direction across the step, related to change of index $\Delta n$. In the current case, a step change of gauge field changes the incident beam's direction abruptly on the interface without lateral shift. On the other hand, at the exit interface that the step change is reversed, the artificial magnetic field also reverses sign to bend light back to its original direction, now with a lateral beam shifting in position due to the change of propagation direction within the medium.

The experimental setup to observe beam shifting can be found in Fig. 3(a). Sixty-one layers of structured PCBs with periodicity 5 mm along $y$ direction is used to construct the bulk sample. The fractal structures are now rotated by 45° to make the working plane being $x' - y$ and foam (in white, with measured relative permittivity 1.002) is used to fill the gaps between the layers. An empirical Gaussian-like microwave beam impinges on the sample and the transmitted beam profile can be measured by a probe antenna mounted on a translational stage. We first consider linear polarized incidence at 12.8GHz with electric field polarized along $z'$ axis, which is a superposition of the two pseudospins $\psi_+$ (with $E_{z'} = H_{z'}$) and $\psi_-$ (with $E_{z'} = -H_{z'}$). Please be reminded that $n_0 = \sqrt{\epsilon_y} \sim 1$. As illustrated in Fig. 3(b), after $E_{z\prime}$ polarized beam (in red) impinges on the sample (in blue), beam shifting occurs as $\psi_+$ ($\psi_-$) will bend upward (downward) in the $y$ direction. Numerical simulation results with $E_{z\prime}$ incidence from left can be found in Fig. 3(c), with a Gaussian beam width of 120 mm to reflect the experimental setup. The position of the bulk gauge field medium is between the two solid lines whose material parameters $\epsilon_x = 1.965, \epsilon_y = 1.016, \epsilon_z = 0.504$ as experimentally retrieved at 12.3 GHz. $H_{z\prime}$ field component emerges after propagating through the bulk sample and the exit $E_{z\prime}$ beam width is slightly widened. Moreover, it can be recognized that the upper exit $H_{z\prime}$ beam has a π phase difference to the lower one. The exit beam profile is a superposition of spatially separated pseudospins: $\psi_+$ and $\psi_-$. $E_{z\prime}$ ($H_{z\prime}$) field will be added up in (out of) phase. A different $\Delta A$ will induce a different beam splitting $L$ and a different exit beam profile of $E_{z\prime}/H_{z\prime}$ component will be induced. In experiment, we measured a line field profile along $y$ direction 30 mm away from the sample exit surface and the experimental results are shown in circles in Fig. 3(d). The $E_{z\prime}$ beam profile without sample at the same position can also be obtained for reference. As discussed previously, the original $E_{z\prime}$ incidence will split into two pseudospins while its beam profile remains. Thus we have the knowledge of beam profiles of the pseudospin beams and their $E_{z\prime}$ and $H_{z\prime}$ components: identical to that of $E_{z\prime}$ beam without sample. As illustrated in the right panel of Fig. 3(b), we can construct the new analytical exit beam profiles at different beam splitting $L$. The best match between the analytical and the experimentally measured beam profiles is with $L$=17.5 mm, where least square method is applied whose results are shown as squares in Fig. 4(d).

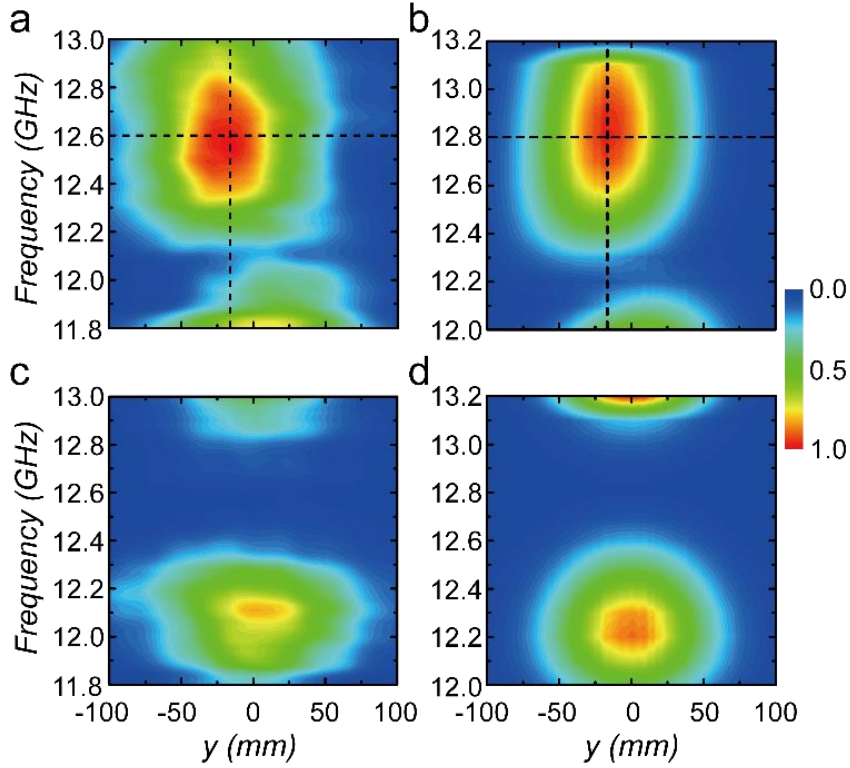

**Fig. 4 Frequency spectrum on beam shift with pseudo spin-up incidence. a-b,** show the measured and theoretical spectrum for co-spin ($\psi_+$) detection. **c-d,** show the measured and theoretical spectrum for the cross-polarization ($\psi_-$) detection.

Now we pursue with dispersive effects. A pseudospin source can be constructed by rotating the horn antenna by 45° from $z'$. $\psi_+$ has its electric field along the $\hat{z}' + \hat{y}$ direction while $\psi_-$ has its electric field along the $\hat{z}' - \hat{y}$ direction. With $\psi_+$ incidence, we measure both the $\psi_+$ and $\psi_-$ field profiles along $y$ direction for different frequencies. The power spectra with normalization to the maximal value of total power ($|\psi_+|^2 + |\psi_-|^2$) for each frequency are shown in Figs. 4(a) and (c). For comparison, Figs. 4(b) and (d) shows the counterpart transmission spectra obtained from simulations and similar behavior in the spectrum is observed. The working frequency is found at about 12.6GHz in experiment (12.8GHz in simulation) where the transmitted beam has smallest $\psi_-$ conversion, as indicated by the horizontal dashed lines. At the working frequency, the transmitted $|\psi_+|^2$ beam shift about 18 mm (17 mm) in the positive $y$ direction from experimental (simulation) results compare to the beam center $y = 0$ without sample, as shown by the vertical dashed lines. At another frequency 12.1GHz in experiment (12.2GHz in simulation), the incident $\psi_+$ beam is largely converted into its cross-polarization $\psi_-$, where the dispersion surfaces are way far away from two circles with two constant decoupled polarizations From the results, we observe that the working frequencies with minimal cross-polarization conversion (and with expected beam shift induced by the gage field) is from around 12.4 to 12.8 GHz, roughly corresponding to the two frequencies satisfying the geometric-mean and harmonic-mean criteria.

In the present work, we have experimentally demonstrated how a spin-dependent 2D gauge field in the real-space can be realized using a metamaterial with biaxial permittivity. Benefitting from the established reduced material parameters approximation, in this work, we relax the requirement of both electric and magnetic responses of an ideal gauge field medium. Only tilted anisotropy from a biaxial permittivity tensor is required, making the realization of a gauge field medium straight-forward. A biaxial metamaterial with fractal geometry in PCB working in the microwave frequency regime is then designed to demonstrate the gauge field. The dispersion surfaces on the working plane are two shifted circles induced by gauge field whose eigen-polarizations are nearly constant as pseudospins, which are defined here as a linear combination of electric and magnetic fields. The beam shifting from a surface pseudo-magnetic field constructed by the designed metamaterial is verified in experiments. We hope that by proposing a simple scheme to achieve gauge field media using biaxial materials, we pave the road to pseudospin optics and future topology related optical components.